# Measurements of a fast nuclear spin dynamics in a single InAs quantum dot with positively charged exciton

X. M. Dou, B. Q. Sun,(a) D. S. Jiang，H. Q. Ni, and Z. C. Niu

*SKLSM, Institute of Semiconductors, CAS, P.O.Box 912, Beijing 100083, China*

**Abstract** – By using highly time-resolved spectroscopy with an alternative $\sigma^+/\sigma^-$ laser pulse modulation technique, we are able to measure the fast buildup and decay times of the dynamical nuclear spin polarization (DNSP) at 5 K for a single InAs quantum dot (QD) with positively charged exciton. It is shown that the nuclear dipole-dipole interaction can efficiently depolarize DNSP with a typical time constant of 500 μs in the absence of external magnetic field. By using an external field of 8 mT to suppress the nuclear dipolar interaction, the decay time turns to be mainly induced by interaction with unpaired electron and extends to about 5 ms. In addition, it is found that the time constant of hole-induced depolarization of nuclear spin is about 112 ms.

**Introduction.** – Hyperfine interaction between the electron in a single quantum dot (QD) and the ensemble nuclear spins recently has attracted considerable attention due to its potential applications in spin-based quantum information processing [1-8]. Dynamical nuclear-spin polarization (DNSP) can be achieved by optically polarized electron in QD even in the absence of external magnetic field provided that the Knight field $B_e$ (~ 1mT) is larger than the local nuclear dipolar field of ~ 0.1mT [9,10]. It is reported that the effective depolarization of DNSP via electron-mediated hyperfine interaction in InGaAs QDs has a decay time of ~ $10^{-3}$ s, being longer than the fast depolarization via nuclear dipole-dipole interaction which is on a time scale of ~ $10^{-4}$ s [1,11]. The absence of nuclear depolarization induced by the nuclear dipolar field in InGaAs QDs was ascribed to the strain-induced quadrupolar interactions, as had been explained qualitatively in InP, InAs and InGaAs QDs, to suppress the effect of nuclear dipolar field [11-13], but a faster depolarization of DNSP with a decay time of 250 μs due to the dipolar interaction was indeed observed in CdSe/ZnSe QDs [14]. So the open question is whether the quadupolar effects have a significant influence to suppress the dipole interaction in InGaAs QDs. The determination of intrinsic time scales of relaxation times is very important for understanding the polarizing and controlling DNSP [13,15,16]. Therefore, it is needed to confirm by the experimental measurements.

In order to obtain the fast nuclear depolarization by nuclear dipolar field experimentally, we have designed a new experimental setup with highly time-resolved spectroscopy. The time resolution of the setup is 100 ns, and such a high time resolution could not be reached by normal "pump-probe photoluminescence" technique as was commonly used in literature [11]. We have measured the buildup and decay times of DNSP in a single InAs QD with positively charged exciton ($X^+$) by using alternative $\sigma^+/\sigma^-$ laser pulse excitation. We show that the optical pulse pumping of the polarized electron can effectively polarize nuclear spins in the time scale of a few hundred microseconds, which is found to be one order of magnitude shorter than the build-up time value reported in Ref. 11. The decay dynamics of DNSP depends drastically on the effect of the applied external magnetic field. It is shown experimentally that in the InAs QD the nuclear dipole-dipole interactions lead to a decay time of ~ 500 μs in the absence of external magnetic field. This decay time turns to be about 5 ms when a small magnetic field of 8 mT is applied in parallel to the sample growth direction, being mainly determined by the electron-nuclear hyperfine interaction. In addition, for the first time, the time constant of the hole-induced depolarization of nuclear spin is measured.

**Experiment.** – The investigated QD samples were grown by molecular beam epitaxy on a semi-insulating GaAs substrate. They consist of, in sequence, an *n*-doped GaAs buffer layer, a 20-period *n*-doped GaAs/Al$_{0.9}$Ga$_{0.1}$As distributed Bragg reflector (DBR), a 2λ GaAs cavity with an InAs QD layer at the cavity antinode, and a top *p*-doped GaAs layer. The ultra-low density of InAs QD layer was formed by depositing nominally 2.35 monolayers (ML) of InAs at a growth rate of 0.001 ML/s. In experiments, the QD sample was mounted in a continuous-flow liquid helium cryostat at 5 K. A mode-locked Ti: sapphire laser with 2 ps pulses and 80 MHz repetition frequency was used to excite the QD sample. The excitation intensity is about 5 $\mu$W. The studied PL emission line of $X^+$ from the positively charged exciton has been well

(a)E-mail: bqsun@semi.ac.cn

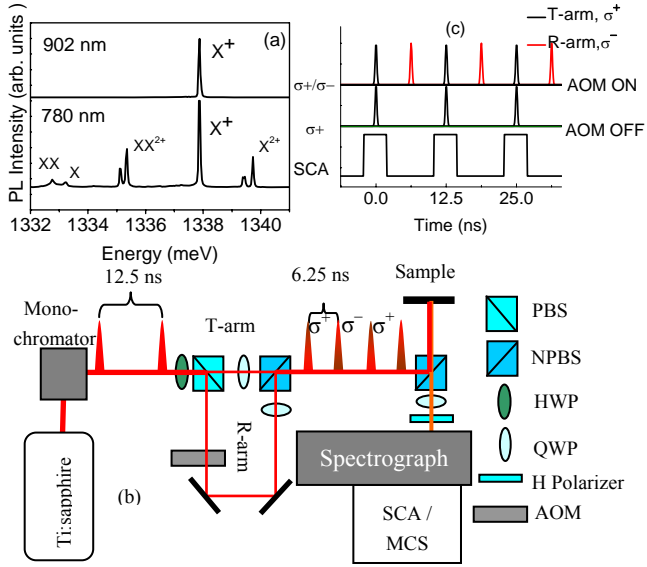

Fig.1: (Color on-line) (a) PL spectra measured at laser excitation wavelengths of 780 and 902 nm, respectively. (b) Schematic diagram of experimental setup of time-resolved PL in which the T-arm and R-arm represent the optical paths of alternating $\sigma^+$ and $\sigma^-$ excitation pulse sequences, separated by 6.25 ns. (c) The corresponding time sequences of laser pulses, modulation by AOM, and a detection window of 4 ns controlled by SCA.

characterized and reported [17]. We note that different charged excitons will exist by exciting the QD wetting layer using a wavelength of 880 nm or exciting GaAs barrier layer using a wavelength of 780 nm. The PL spectra obtained under the latter excitation conditions are shown in the lower half of Fig. 1(a). In order to generate only a singly charged exciton in QD, the exciting laser wavelength is intentionally tuned to 902 nm to excite the QD sample by the GaAs LO-phonon-assisted resonance. The PL result is shown in the upper half of Fig. 1(a). The emitted luminescence was collected by an objective (NA: 0.5), spectrally filtered by a 0.5 m monochromator, and then detected by a silicon charge coupled device (CCD) and an avalanche photodiode (APD). For the polarization PL measurements, the excitation pulses were circularly polarized ($\sigma^+$) using a $\lambda/4$ wave plate. The luminescence emission was analyzed by a $\lambda/4$ wave plate and a linear polarizer to distinguish different polarization components.

Note that the polarization degree of the PL emission originated from localized electronic states is particularly sensitive to the variations of the nuclear spin polarization. It is due to the fact that the nuclear polarization which induces an effective field as so-called Overhauser field has an influence as an retro-action on the electron spin orientation [5,18] As has been reported [5] that the value of integrated PL circular polarization degree ($P_c$) obtained under randomly oriented nuclear spins is smaller than that under polarized ones, the former is only about 40% of the latter. When the positively charged exciton $X^+$ is created in a single QD, a trion composed of two holes and one electron will exist, where the two holes form a spin singlet, and the unpaired single electron interacts with the nuclei during the radiative lifetime of the excitonic recombination in about 1 ns. The PL measurements of the circular polarization of the $X^+$ emission in QDs following circularly polarized laser excitation in such a time scale thus will directly probe the spin polarization of the electron as $\langle \hat{S}_z^e \rangle = -P_c/2$. In order to make an in-depth study on the dynamics of nuclear spins in single QD, a so-called $\sigma^+/\sigma^-$ modulation technique is employed to excite the sample, where two pulse sequences, i.e. $\sigma^+$ and $\sigma^+/\sigma^-$ sequences, can be realized with alternating $\sigma^+$ and $\sigma^-$ polarizations, separated by 6.25 ns. The optical path used in the measurements is indicated in Fig. 1(b). An acousto-optical modulator (AOM) is placed on the R-arm, serving a fast switching between $\sigma^+/\sigma^-$ and $\sigma^+$ excitation conditions, with rise- and fall times of less than 100 ns. The alternating $\sigma^+/\sigma^-$ pulse sequences excite the sample as the AOM switches on, whereas only $\sigma^+$ pulse sequences work if the AOM switches off. The time-dependent change of the PL polarization is measured by setting a detection window of 4 ns using a single channel analyzer (SCA) after the QD is excited by the $\sigma^+$ pulses, as indicated in Fig.1 (c). In order to measure with the different time scales of DNSP, the driven frequencies of the AOM is chosen as a 200 Hz or 20 Hz，and the driven amplitude is taken as 2 V. The nuclear spins are polarized during the $\sigma^+$ pulse sequences excitation of the QD sample, and then nuclear spins will be depolarized during $\sigma^+/\sigma^-$ pulse excitation when AOM is on. These DNSP processes will imprint on the variations of PL polarization, demonstrating the buildup and decay times of DNSP following AOM switching off and on, respectively. The output of SCA is input into a multi-channel scaler (MCS) with a measuring range of 10 ms (100 ms), and each time bin is 10 μs (100 μs), respectively. Such a high time-resolution of the setup enables us to measure the fast dynamics of nuclear spins when it occurs. In fact, this experimental setup is similar to the so-called "pump-probe PL" technique [11], but has a better time-resolution of 100 ns which is limited by dwell time of MCS and the rise- and fall times of AOM.

**Results and discussion.** – Using the $\sigma^+/\sigma^-$ modulation technique, we have first measured the PL intensity of $\sigma^+$ and $\sigma^-$ components as a function of time after the QD sample is excited by the $\sigma^+$ pulses, shown as ($\sigma^+, \sigma^+$) and ($\sigma^+, \sigma^-$) in Fig.2 (a) and (b), respectively. Here it is noted that PL value measured at time zero is the same as one measured at time 5 ms which corresponds to one modulation period of AOM. As expected, at every 2.5 ms interval, following the AOM switching off and on, the $\sigma^+$ and $\sigma^-$ PL components have an oppositely-directed variation with the time. This result reflects the variation of nuclear field (Overhauser field) acting on the electron spin orientation. From these curves we can obtain the time-dependent circular polarization degree ($P_c$) according to the expression $P_c = (I_{\sigma+} - I_{\sigma-})/(I_{\sigma+} + I_{\sigma-})$, where $I_{\sigma+}$ and $I_{\sigma-}$ are the PL emission intensities of each time bin in MCS with $\sigma^+$ and $\sigma^-$ components. The derived $P_c$ data are shown in Fig. 2(c) by open circles, where the uprising and decaying parts of the obtained curve correspond to the buildup and decay processes of DNSP, respectively. From this curve, the obtained buildup and decay times are 270 μs and 500 μs, respectively, by a single exponential fit to the data[11] as indicated by two different red lines in the figure. In fact, the same order of decay time ~ 250 μs was reported in CdSe/ZnSe QD and attributed to the nuclear spin dipole-dipole interaction [14]. In general, a typical time constant of such interaction is expected to be $10^{-4}$ s, or a few hundreds of microsecond, just as what we have experimentally obtained [1,18]. However, it is noted that for the negatively charged exciton $X^-$ in InGaAs QDs, the corresponding decay time was reported to be as large as 1.9 ms [11], which is ascribed to the hyperfine interaction between the residual electron and nuclei after the exciton $X^-$ recombination with a lifetime of ~ 1ns. This value of decay time is actually about one order of magnitude larger than our result reported here.

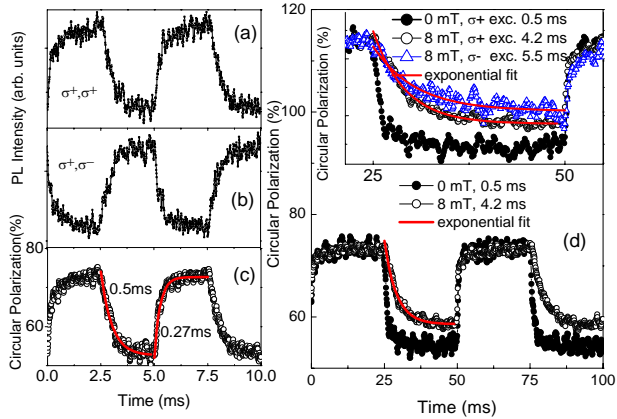

Fig.2: (Color on-line) Time-resolved PL intensity of $X^+$ emission for the components of ($\sigma^+$, $\sigma^+$) and ($\sigma^+$, $\sigma^-$) in (a) and (b), respectively. (c) Time-dependent circular polarization $P_c$ modulated by AOM with a modulation period of 5 ms. (d) Time-dependent circular polarization $P_c$ in the absence of $B_{ext}$ field (shown as solid circles) and $B_{ext} = 8$ mT (shown as open circles). Here the modulation period of AOM is 50 ms. In the top inset: The circular polarization under the condition of $B_{ext} = 0$ (solid circles) and $B_{ext} = 8$ mT. Two constant helicities $\sigma^+$ (open circles) and $\sigma^-$ (open triangles) are used to excite the QD sample in T-arm.

Now we apply a small magnetic field $B_{ext}$ (~8 mT) in parallel to the sample growth direction, the strength of $B_{ext}$ is larger than the dipolar field of ~ 0.1mT, but is still much smaller than the dispersion of the nuclear hyperfine field of ~ 26 mT in single QD [5]. The applied $B_{ext}$ will ensure that the depolarization of DNSP via dipolar interaction between nuclear spins is suppressed [9,10] However, the $B_{ext}$ field is still too small to have any significant influence on the electron spin polarization. Thus the measured PL circular polarization, i.e., electron spin polarization, will follow the variation of nuclear field, i.e. the depolarization of DNSP, due to the fact that electron-nuclear hyperfine interaction will be responsible for the measured change in circular polarization. Such a relaxation process of nuclear spin is indeed observed in our experiment when applied $B_{ext} = 8$ mT, as shown in Fig.2 (d) by open circles, where the data at $B_{ext} = 0$ (solid circles) are also shown in the figure for comparison. From the curve in Fig.2 (d) the derived decay time is 4.2 ms at $B_{ext} = 8$ mT by using a single exponential fit to the data. The value is about one order of magnitude longer than the case of $B_{ext}=0$ where the decay time is 0.5 ms and the relaxation is dominated by nuclear dipolar interaction. It is noted that a decay time of the same order due to the depolarization of DNSP has been reported previously by the other group [11], and which is also ascribed to the electron-nuclear hyperfine interaction. In addition, it is interesting to note that in Fig.2 (d) all the rising edges of different curves measured at AOM switching off, even the applied magnetic field conditions are different, are overlapped with each other very well. It means that the buildup time of DNSP is independent of the $B_{ext}$ field when the strength of $B_{ext}$ is within the range of a few mT. This means that the polarized electron is responsible for the process of DNSP, and the nuclear dipole-dipole interaction is insignificant in the building-up process of the nuclear polarization. The conclusion is further confirmed by measuring the buildup and decay times of DNSP when the $\sigma^+$ excitation pulses are changed to be $\sigma^-$ ones, as shown in the top inset of Fig.2(d), where the decay times are 4.2 and 5.5 ms for either $\sigma^+$ or $\sigma^-$ excitation pulses in T-arm, respectively, at a fixed $B_{ext}$ field. It is found that both time constants are very close to each other and can be taken as nearly the same. In fact, the obtained buildup time is always ~ 270 μs, independent of three different experimental conditions shown in Fig. 2 (d). Thus, it is concluded that the electron-nuclear spin flip rate which is dominant in the buildup process is not perturbed by small $B_{ext}$ field in these measurements. Instead, the small $B_{ext}$ field really can efficiently suppress the nuclear dipolar interaction.

In the above-mentioned measurements, the decay time of the polarized nuclear spins is detected based on the alternative $\sigma^+/\sigma^-$ laser pulse modulation technique, and the spin of QD electron changes up and down alternately, where the lifetime of electron is ~1 ns and modulation period of laser is 6.25 ns. When the mechanism of the depolarization of DNSP is the hyperfine interaction between the electron and the nuclear spins, it is given by[19]

$$\hat{H}_{hf} = \frac{v_0}{2}\sum_j A^j |\psi(R_j)|^2 (\hat{I}_z^j \hat{S}_z + \frac{\hat{I}_+^j \hat{S}_- + \hat{I}_-^j \hat{S}_+}{2}) \qquad (1)$$

where $v_0$ is the volume of the unit cell, $A^j$ is the constant of the hyperfine interaction, $|\psi(R_j)|^2$ is the electron density at location $R_j$ of the $j$th nuclear spin, $\hat{S}$ and $\hat{I}^j$ are the electron and nuclear-spin operators, respectively. The sum goes over all nuclei. Eq. (1) can be decomposed into two parts: A static part ($\propto I_z^j S_z$), affecting the energies of the electron and the nuclear spins, and a dynamical part ($\propto \sum_i [\hat{S}_+ \hat{I}_-^i + \hat{S}_- \hat{I}_+^i]$), allowing for the transfer of angular momentum between the two spin systems. The latter term is an important term to describe the dynamics of DNSP. If the initial nuclear spins have a randomly oriented distribution, the electron-nuclear spin flip-flop rates are the same for two different spin directions, i.e., the sum $\sum_i \hat{S}_+ \hat{I}_-^i = \sum_i \hat{S}_- \hat{I}_+^i$. This means that the $\sigma^+/\sigma^-$ modulation pulses do not perturb the nuclear spin distribution [5]. However, if the nuclear spins are polarized, for example, the number of polarized QD nuclear spins with spin operator $\hat{I}_-^i$ or $\hat{I}_+^i$ is about 10% in our case [20,21], then the polarized nuclear spins will be depolarized by electron-nuclear spin flip-flop via a term of $\sum_i \hat{S}_+ \hat{I}_-^i$ or $\sum_i \hat{S}_- \hat{I}_+^i$, whereas other un-polarized nuclei is kept in place. At the end, the polarized nuclear spins (nonequilibrium nuclear spins) will relax into the equilibrium of the nuclear spins via electron-nuclear spin flip-flop, as we have observed in the experiment.

It is noted that a typical nuclear-spin decay time ($T_{1e}$) by hyperfine interaction can be estimated by [7, 22, 23]

$$\frac{1}{T_{1e}} = (\frac{\bar{A}_e}{N\hbar})^2 \frac{2 f_e \tau_c}{1+(\Omega_e \tau_c)^2} \qquad (2)$$

where $\bar{A}$ is the average of the hyperfine constants $A^j$, $N$ is the number of relevant nuclei in QD, $f_e$ is a fraction of time the QD contains an unpaired electron, $\tau_c$ is the electron spin correlation time, $\Omega_e = \Delta E_e^z / \hbar$ is the electron angular Larmor frequency, and $\Delta E_e^z$ is the electron Zeeman splitting. The following values are assumed for the estimation of nuclear-spin decay time ($T_{1e}$): $\bar{A}_e$ = 51.5 $\mu$eV ($A_{As}$ = 47 $\mu$eV and $A_{In}$ = 56 $\mu$eV ) [5], $N \sim 10^4$ the number of polarized nuclei [21], $\Omega_e \sim$ 11.6 GHz [24], $\tau_c \sim$ 50 ps [22,23], and $f_e \sim$1/12 taking into account that the lifetime of electron is ~1ns, and the laser pulse repetition rate is 80 MHz (a period of 12.5 ns). Then we obtain $T_{1e}$ = 2 ms from Eq. (2), which value is reasonably close to the measured decay time of about 5 ms.

For a hole in QD, on the other aspect, the Fermi contact coupling is expected to be much weaker because of the *p* symmetry of the valence band states [18]. Thus the hyperfine interaction between it and nuclear spins will be much smaller. However, theoretical and experimental studies have shown recently that such interaction is stronger than the previously expected one due to the band hybridization [2,4,25]. The corresponding hyperfine constants between the hole and nuclei of As and In are reported to be $C_{As}$ = 4.4 $\mu$eV and $C_{In}$ = 4.0 $\mu$eV, respectively [4]. These constants are larger than what are expected before considering the band hybridization effect, but are still relatively small, being about one order of magnitude smaller than the hyperfine coupling of electron with nuclear spins. Thus, if the hole correlation time and Zeeman splitting are assumed to be on the same order as those of electrons in the QD, and taking $f_h \sim$1 in Eq. (2), where $f_e$ is replaced by $f_h$, i.e. the fraction of time the QD contains an residual hole. Then the modified Eq. (2) predicts that the hole-related decay time, i.e. the decay time induced by the hyperfine interaction between residual hole and nuclear spins after excitonic recombination of exciton $X^+$, is in the range of hundred of millisecond.

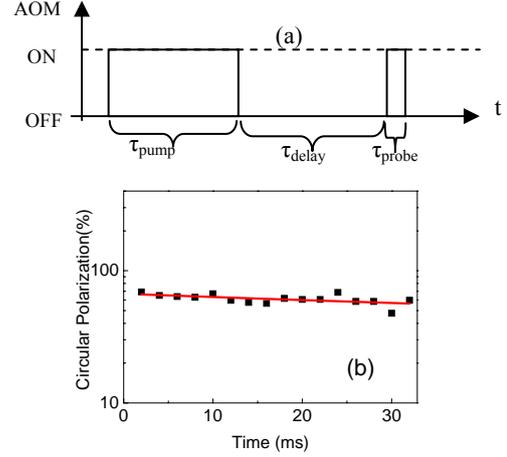

Fig.3: (Color on-line) (a) Schematic diagram of the "pump-probe PL" setup, where only the $\sigma^+$ pulse sequences of the T-arm are used. The $\tau_{pump}$ and $\tau_{probe}$ are 10 ms and 40 $\mu$s, respectively, controlled by AOM. $\tau_{delay}$ is variable by steps with each step of 2 ms. (b) Circular polarization $P_c$ versus delay time.

In fact, this estimated decay time is about two orders of magnitude longer than that of the electron-induced nuclear decay time. In order to experimentally check the hole-induced time constant, the "pump-probe PL" setup is employed for the measurement, where the time durations of $\tau_{pump}$, $\tau_{delay}$, and $\tau_{probe}$ are controlled by switching-on and switching-off of AOM, respectively, as schematically shown in Fig. 3(a). After a pumping pulse of $\tau_{pump}$, the time-dependent polarization PL is measured by the probe pulse of $\tau_{probe}$ after a varying and sufficiently long delay time $\tau_{delay}$ (> 2 ms). In addition, the chosen length of $\tau_{probe}$ (40 $\mu$s) is far less than the above-measured electron-mediated nuclear spin decay time $\tau_{decay}$. Figure 3 (b) shows the measured circular polarization of $P_c$ as a function of the delay time of probe pulse, $\tau_{delay}$. The characterization time of depolarization of DNSP is found to be 112 $\pm$ 41 ms by fitting a single exponential to the data as indicated by the red line. The result is quite consistent with the value estimated by Eq. (2) in the order of magnitude.

A typical time scale of DNSP decay induced by the nuclear dipole-dipole interaction is $\sim 10^{-4}$ s, which is one order of magnitude shorter than the reported nuclear spin depolarization induced by electron-mediated hyperfine interaction in QD. Therefore, it is expected that the nuclear dipolar interaction, instead of hyperfine interaction, may more efficiently depolarize the nuclear spins. This expected result has been indeed observed in our experiment when no magnetic field is applied. A typical time constant of 500 $\mu$s is obtained. It is noted that, using the similar $\sigma^+/\sigma^-$ modulation technique, a faster depolarization of DNSP with a similar decay time of 250 $\mu$s was observed in CdSe/ZnSe QDs due to the dipolar interaction.[14] It is thought that perhaps a reasonably high time-resolution of the experimental setup may be important in order to observe the faster dynamics of DNSP. We also note that it was considered that quadupolar effect in InP, InGaAs and InAs quantum dots due to both alloying and strain would suppress the dipolar interaction [11-13] which might result in a relaxation time longer than what expected from the dipole interaction. However, this effect was not explicitly observed in our measured results. Thus, a more detailed analysis of the sample condition, or a further theoretical work, is needed to clarify the related physical processes.

**Conclusion.** – In conclusion, by designing a new experimental setup with highly time-resolved spectroscopy, the fast buildup and decay times of DNSP have been measured. It is shown that the optically-pumped polarized electron can effectively polarize nuclear spins in a buildup time as short as a few hundred microseconds. The nuclear dipolar interaction can efficiently depolarize DNSP with a typical constant of 500 μs in the absence of the external magnetic field. By applying $B_{ext}$ field to suppress the nuclear dipolar interaction, the decay time is turned to be about 5 ms induced by QD electrons. The relaxation time of hole-mediated depolarization of DNSP is found to be as long as 112 ms by "pump-probe PL" technique.
.

***

This work is supported by the National Basic Research Program of China and NSPC under Grant Nos. 2007CB924904 and 90921015.